\documentclass[11pt]{article}

\usepackage[final]{acl}

\usepackage{times}
\usepackage{latexsym}

\usepackage[T1]{fontenc}
\usepackage[utf8]{inputenc}

\usepackage{microtype}

\usepackage{inconsolata}

\usepackage{graphicx}
\usepackage{booktabs}
\usepackage{amsmath}
\usepackage{tikz}
\usetikzlibrary{shapes.geometric, arrows.meta, positioning, fit, calc, backgrounds}
\usepackage{pgfplots}
\pgfplotsset{compat=1.17}
\usepackage{multirow}
\usepackage{xcolor}
\usepackage{enumitem}
\usepackage{subcaption}
\usepackage{array}
\usepackage{float}
\usepackage{multirow}
\usepackage[most]{tcolorbox}
\usepackage{svg}

\usepackage{capt-of}

\usepackage{framed}
\newenvironment{challengebox}
{%
  \setlength{\fboxsep}{6pt}%
  \MakeFramed{\advance\hsize-\width\FrameRestore}%
  \noindent\small\normalfont\ignorespaces
}
{%
  \par\endMakeFramed
}

\usepackage{colortbl}
\definecolor{EcoHighlight}{RGB}{255,228,228}
\colorlet{generate}{EcoHighlight}

\providecommand{\ecomax}[1]{%
  \begingroup
  \setlength{\fboxsep}{0pt}%
  \smash{\colorbox{EcoHighlight}{%
    \makebox[2.2em][c]{\strut #1}%
  }}%
  \endgroup
}
\title{ReconPotBench: Evaluating LLM Honeypots in Multi-Round Linux Reconnaissance}

\author{Qianli Wang$^{1,2}$ ~~~ Junchao Wu$^{1}$ ~~~ Binkai Ou$^{3}$ ~~~ Lidia S. Chao$^{1}$ ~~~ Derek F. Wong$^{1}$\thanks{Corresponding author.}\\
$^{1}$NLP$^2$CT Lab, Faculty of Information Science and Computing, University of Macau\\
$^{2}$School of Computer Science and Technology, Shandong University\\
$^{3}$BoardWare Information System Limited\\
\texttt{wqlinfosec@gmail.com} ~~~ 
\texttt{nlp2ct.junchao@gmail.com} ~~~ \texttt{\{derekfw,lidiasc\}@um.edu.mo}
}

\begin{document}
\maketitle

% \begingroup
%   \renewcommand{\thefootnote}{\fnsymbol{footnote}}
%   \footnotetext[2]{Corresponding Author}
% \endgroup

\begin{abstract}
Large language models (LLMs) are increasingly used to build interactive honeypots, yet existing evaluations mainly assess isolated command outputs and cannot determine whether a system remains believable when later commands depend on earlier results. We present ReconPotBench, a multi-round benchmark for Linux reconnaissance honeypots. It contains 1{,}032 three-round reconnaissance interactions across eight reconnaissance categories, with reference outputs collected from a real Linux environment. Under a unified testing protocol, we compare Cowrie, open-source LLMs, and closed-source LLMs along three dimensions: round-level authenticity, cross-round contextual consistency, and hacker-oriented reconnaissance value. Results show that multi-round evaluation exposes failures hidden by single-round testing, as authenticity declines with interaction depth and both open-source and closed-source LLMs struggle to maintain coherent state across rounds. LLM-based honeypots nevertheless outperform Cowrie in authenticity and reconnaissance value, while Cowrie remains more consistent because of its rigid templated behavior. Category-level analysis further shows that tasks requiring systems to preserve network relationships across rounds are harder than tasks centered on explicit entities such as ports. ReconPotBench enables reproducible evaluation of LLM honeypots across authenticity, entity consistency, and hacker-perspective utility in dependent multi-round interactions. The dataset is available at: \url{https://github.com/qianli-wang/ReconPotBench}.
\end{abstract}

\section{Introduction}

\label{sec:intro}

Reconnaissance is the critical first phase of the cyber attack lifecycle, and it forms the basis for all subsequent attack activities. As specified in the MITRE ATT\&CK Framework~\citep{mitre2025reconnaissance1}, reconnaissance is the initial tactic adopted by adversaries to collect information for targeting operations. Studies have shown that attackers spend most of their time on this phase~\citep{reconnaissance_time}. Using large language models (LLMs) as an interactive deception mechanism to counter reconnaissance activities is a promising paradigm in active defense. Evaluating the effectiveness of these LLM-based defense methods is critical, since effective detection and analysis in the reconnaissance phase allow defenders to identify and neutralize threats before they develop into active exploitation~\citep{LLM_in_the_Shell,HoneyLLM,LLM_Honeypot,LLMHoney,HoneyGPT}.

\begin{figure}[!t]
    \centering
    \includegraphics[width=\linewidth]{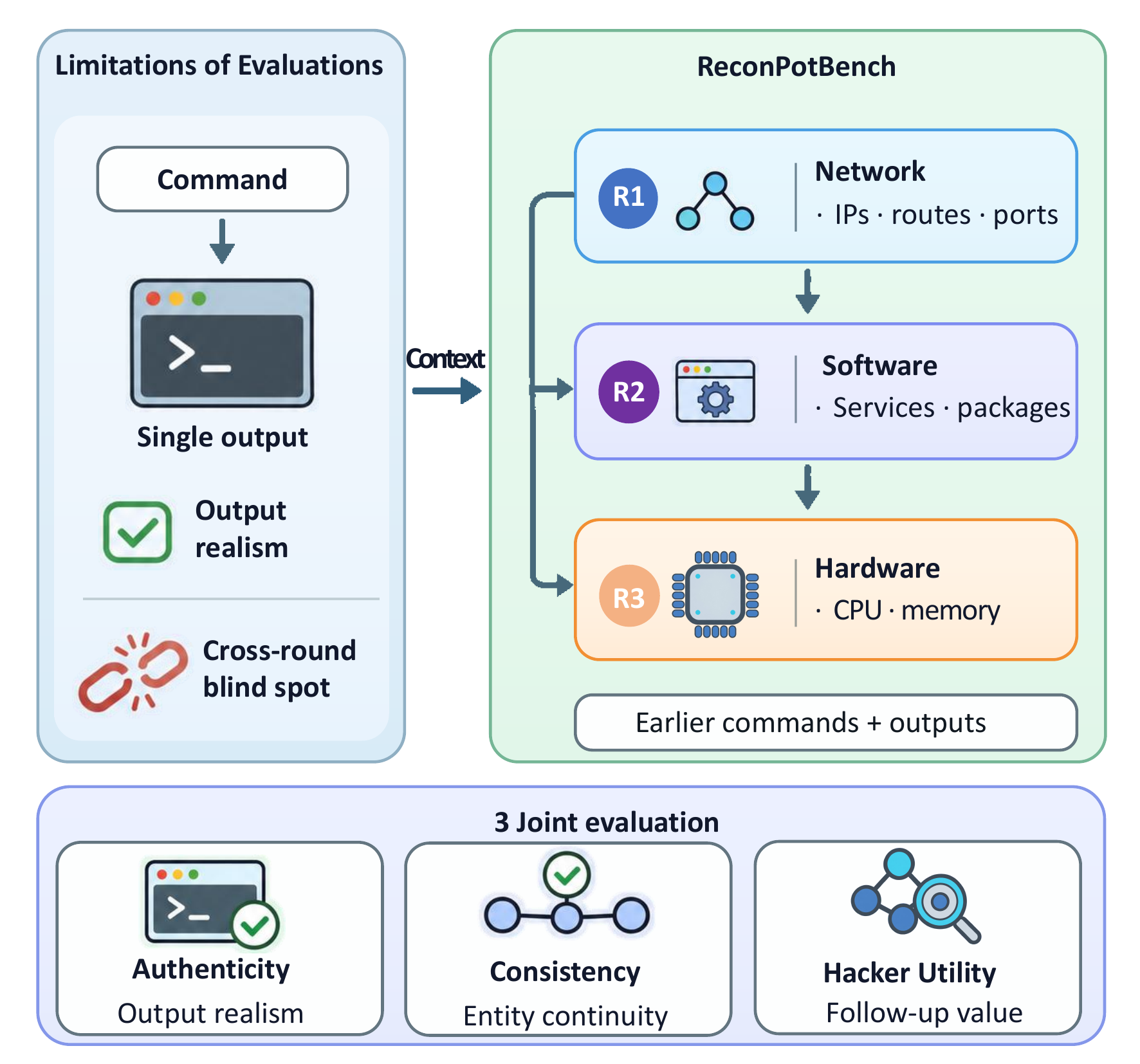}
    \caption{Motivation of our benchmark: existing Linux command-output datasets mainly focus on isolated single-round interactions, while realistic reconnaissance involves multi-round command sequences with contextual dependencies across rounds.}
    \label{fig:background}
\end{figure}

However, existing evaluations of LLM-based honeypots are not designed for multi-round Linux reconnaissance. Prior work mainly falls into three lines: feasibility studies showing that LLMs can simulate shell interactions~\citep{Chatbots}; per-response realism evaluations based on expert judgment, template-based references, or surface-level similarity metrics~\citep{LLM_in_the_Shell,LLM_Honeypot,LLMHoney}; and longer interactive evaluations, attacker engagement, or session-level deception quality~\citep{HoneyLLM,HoneyGPT,ShellBox}. These studies demonstrate the promise of LLM honeypots, but their evaluations miss the dimensions attackers would actually notice during Linux reconnaissance. In realistic Linux reconnaissance, attackers assess not only whether each output looks authentic, but also whether entities such as IP addresses remain consistent across rounds, and whether the interaction continues to support meaningful follow-up probing. Failures in authenticity, contextual consistency, or hacker-perspective utility can reveal the deception, yet prior evaluations do not measure these dimensions jointly in sequentially dependent command-output interactions.

Figure~\ref{fig:background} highlights this mismatch between existing single-round evaluation settings and the context-dependent nature of realistic reconnaissance behavior, motivating the need for a benchmark that can evaluate both round-level realism and cross-round contextual coherence. 
We introduce ReconPotBench, a benchmark of fixed three-round Linux reconnaissance interactions that jointly evaluates round-level authenticity, cross-round consistency, and attacker-perspective utility. Our approach stems from the observation that real-world attackers do not execute commands in isolation. They instead carry out extended interaction sequences, where each command relies on information gathered from prior interactions. This establishes multi-round testing as a necessary component rather than an optional step. Single-round evaluation only verifies whether an individual response appears authentic. It cannot determine whether the system preserves a coherent host state across turns, nor whether the accumulated outputs provide valuable information for follow-up reconnaissance. This distinction is important because attackers judge a system not only by local output realism, but also by whether entities remain consistent, and whether the interaction continues to support meaningful probing. 

To evaluate these properties, we build a dataset with 1{,}032 three-round interaction test cases across eight reconnaissance scenarios. Each test case follows a Network~$\rightarrow$~Software~$\rightarrow$~Hardware hierarchy, with ground-truth outputs collected from a physical Linux system. The commands are constructed from standard Linux command options~\citep{Command_line} to ensure valid reconnaissance intent. This design enables ReconPotBench to measure three properties: round-level authenticity, cross-round contextual consistency, and hacker-perspective utility. Together, they capture not only whether each output looks realistic, but also whether the full multi-round interaction remains coherent and useful from an attacker’s perspective.

\noindent{We summarize our main contributions as follows:}
\begin{itemize}
\item We introduce ReconPotBench, a multi-round benchmark for evaluating LLM-based honeypots in Linux reconnaissance. Instead of checking isolated command responses, ReconPotBench turns reconnaissance into 1{,}032 three-round Network$\rightarrow$Software$\rightarrow$Hardware interaction chains with real-system reference outputs.

\item We develop a testing protocol around three hacker-relevant dimensions: round-level authenticity, contextual consistency, and hacker-perspective utility. Applying it to Cowrie, closed-source LLMs, and open-source LLMs shows why single-round evaluation is not enough: LLM honeypots can produce more realistic and more valuable reconnaissance outputs than Cowrie, but their authenticity drops as the interaction deepens and their cross-round entity consistency remains weak.

\item We conduct a category-level evaluation over eight reconnaissance categories. The results show that scenarios with explicit entities, such as ports or service names, are easier to maintain, while tasks involving implicit relations, such as routing logic, DNS configuration, and interface-address bindings, expose more state-tracking failures. This points to a practical bottleneck for LLM honeypots: plausible terminal text alone is not enough; the system also needs stable host-state memory across rounds.
\end{itemize}

\section{Background and Related Work}
\label{sec:related}

Recent studies have begun to use LLMs as response generators for shell or terminal honeypots. The evaluation question has therefore shifted from whether a model can answer a command to whether the honeypot remains believable as an attacker keeps interacting with it. Existing work approaches this problem from three angles: feasibility, single-response realism, and longer interactive behavior.

Early work first establishes feasibility. ~\citet{Chatbots} shows that chatbot-based systems can emulate terminal interactions across operating systems and common application and network tools, suggesting that LLMs can serve as honeypot interfaces. This line of work mainly answers whether such interfaces can be built, rather than how closely each generated command output follows real shell behavior.

A second line evaluates command-level realism. ~\citet{LLM_in_the_Shell} asks experts to judge generated responses against standard Linux shell behavior on a per-command basis. ~\citet{LLM_Honeypot} compares generated outputs with Cowrie references and further validates the system through deployment as an interactive SSH honeypot. ~\citet{LLMHoney} tests 138 representative Linux commands against real shell outputs, reporting automatic quality metrics together with latency and memory cost. However, their limitation is also clear: each command is treated as a standalone case, so contradictions that only appear after several dependent rounds are largely outside the scope.

A third line studies longer sessions. ~\citet{HoneyLLM} moves beyond single-command realism and identifies accuracy and consistency as key challenges for natural shell emulation, using prompt engineering to support longer interactions. ~\citet{HoneyGPT} compares a ChatGPT-powered terminal honeypot with traditional honeypots and real systems, and also reports a three-month deployment alongside Cowrie to observe attacker engagement in practice. ~\citet{ShellBox} targets multi-round inconsistency directly through interaction history pruning and dynamic error simulation. These studies show that interaction-level behavior matters, but their evaluations are still centered on output similarity or coherence. They do not directly evaluate whether a dependent multi-round reconnaissance chain preserves hacker-relevant state and remains useful for follow-up probing.

ReconPotBench is designed as a direct response to the evaluation problems left by these three lines of work. For feasibility-oriented studies, the question is no longer whether an LLM can imitate a shell at all, but whether such a shell can be tested under realistic Linux reconnaissance; we therefore fix the benchmark around concrete terminal commands and real host outputs. For single-command realism studies, the missing piece is that a locally plausible response may still fail once later rounds depend on earlier rounds; we therefore organize each case as a controlled three-round Network$\rightarrow$Software$\rightarrow$Hardware reconnaissance chain and evaluate round-level authenticity, cross-round consistency, and hacker-perspective utility together. For longer-session studies, interaction quality or coherence alone still does not show whether the session is valuable from an attacker's viewpoint; we therefore add hacker-perspective utility to measure whether the accumulated outputs support meaningful follow-up probing. This design makes ReconPotBench more than another shell-emulation test: it measures output authenticity, cross-round entity consistency, and the value of outputs for supporting hackers' follow-up probing across dependent reconnaissance rounds.

\section{Methodology}

\label{sec:method}

\subsection{Benchmark Architecture}

\begin{figure*}[t]
    \centering
    \includegraphics[width=\textwidth]{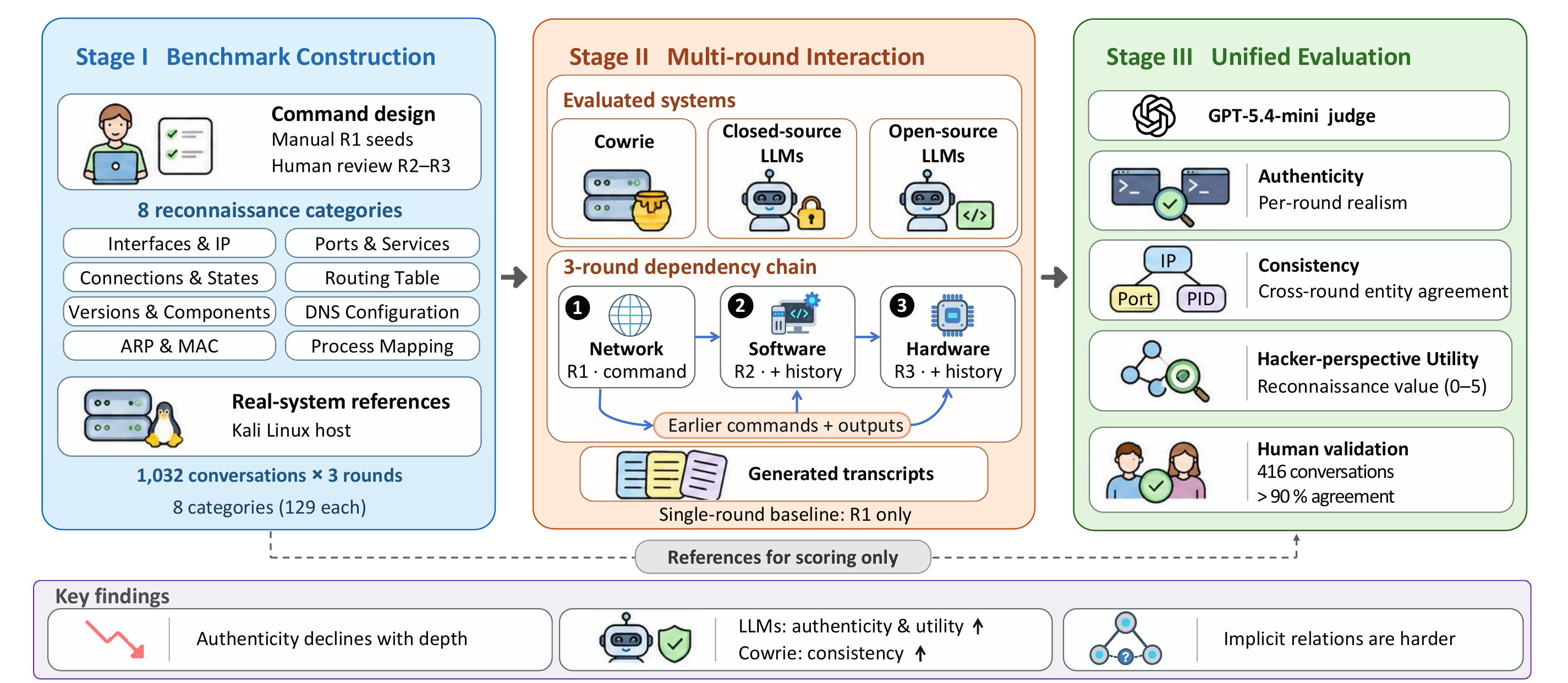}
    \caption{Overview of ReconPotBench. The benchmark targets multi-round Linux reconnaissance interactions, in which later commands depend on earlier context, and enables unified evaluation of systems along three dimensions: round-level authenticity, cross-round consistency, and hacker-perspective utility.}
    \label{fig:overview}
\end{figure*}

Figure~\ref{fig:overview} presents the overall design of ReconPotBench. Each instance is a three-round Linux reconnaissance chain that keeps a single reconnaissance objective across rounds, allowing us to test whether a system can remain believable as later commands depend on earlier outputs. We construct 1{,}032 multi-round interactions from a real Linux environment and organize them into eight representative reconnaissance scenarios. Under the unified protocol, the evaluated system receives only the current command in Round~1, while Rounds~2 and~3 additionally include the previous commands and generated outputs from the same interaction. The real-system reference outputs are never exposed during generation and are used only for evaluation. ReconPotBench compares traditional honeypots, closed-source LLMs, and open-source LLMs along three dimensions: round-level authenticity, cross-round consistency, and hacker-perspective utility. This design enables controlled comparison of whether a system can generate realistic terminal outputs, preserve coherent host state, and provide valuable reconnaissance information throughout a dependent multi-round interaction.

\subsection{Dataset Construction}
\label{sec:dataset}

\subsubsection{Three-Round Command Hierarchy}

Our dataset follows a Network Layer~$\rightarrow$~Software Layer~$\rightarrow$~Hardware Layer hierarchy to model dependent Linux reconnaissance. Inspired by MITRE ATT\&CK TA0043~\citep{mitre2025reconnaissance1} and common attack workflows~\citep{reconnaissance2}, this design reflects how attackers first gather network-level clues and then use them to inspect related services, software, and host resources. Round~1 probes IPs, routes, ports, and interfaces; Round~2 examines services, packages, configurations, and versions; and Round~3 probes hardware or host-resource context. This structure turns each instance into a connected reconnaissance chain, enabling ReconPotBench to evaluate authenticity, cross-round contextual consistency, and hacker-relevant reconnaissance value within the same interaction. The hierarchy provides a controlled sequence for examining related information within a case; it does not prescribe a universal order for real reconnaissance. For example, an interface-centered case can examine its addresses, kernel network configuration, and driver and bus location. MITRE reconnaissance and discovery behaviors inform task selection, while the fixed three-round structure is a benchmark design choice.

\subsubsection{Eight-Category Command Taxonomy}
\label{sec:taxonomy}

The eight categories distinguish the objects inspected and the information returned, whereas the three-layer structure specifies how inspection proceeds within each case. For example, listening-port inspection asks which services are listening, connection-state inspection asks which communications are active, and connection-to-process mapping asks which processes are responsible for those communications. These tasks require different information and entity relationships, so we evaluate them separately. The categories cover the common inspection tasks selected for this benchmark and are not an exhaustive taxonomy of reconnaissance. We include 129 cases per category to provide balanced coverage, rather than to represent their frequency in real attacks. Table~\ref{tab:categories} summarizes the categories, and Appendix~\ref{Reconnaissance_Categories} provides the reconnaissance references and representative commands for all eight categories.

\subsubsection{Data Generation Methodology}

\textbf{First-Round Command.} We manually design 1{,}032 first-round commands, with 129 commands per category. This design is important because the first command determines how the three-round dependency chain begins. It defines the initial reconnaissance intent, controls category coverage, provides the basis for meaningful follow-up commands, and helps ensure that subsequent evaluations are valid. Each command is derived from the eight reconnaissance scenarios in Section~\ref{sec:taxonomy}, instantiated using standard Linux command options from authoritative documentation~\citep{Command_line} and penetration-testing practice, and validated on a Linux system before inclusion.

\textbf{Second-Round \& Third-Round Commands.} We generate the next two rounds of commands using GPT-5.4 with a prompt described in Appendix~\ref{para:appendix_multiround_prompt}. Because the first-round commands are manually designed and category-aligned, the LLM is used only to extend a controlled reconnaissance chain rather than to define the task from scratch. Given a first-round command, the prompt identifies its scenario and generates one software-layer command and one hardware-layer command under the Network Layer~$\rightarrow$~Software Layer~$\rightarrow$~Hardware Layer progression~\citep{reconnaissance2}. Both follow-up commands must remain within the same reconnaissance scenario and preserve logical dependence on prior rounds. We manually review all generated commands for technical correctness and contextual suitability.

\subsubsection{Ground Truth Collection and Experimental Validation}

We use a physical Linux host as a controlled, target-like reference environment for collecting real terminal outputs. All commands are executed on this host, and the resulting outputs are used as reference labels. Real-system labels are necessary because authenticity cannot be reliably judged against simulated or template-based outputs. We deploy standard penetration-testing tools and services on the host to support realistic Linux reconnaissance, preserve formatting and special characters in plain-text outputs, and reset system state when needed to avoid interference across multi-round command chains. We manually inspect the collected outputs to ensure that all commands execute correctly and that the collected references are free from execution or collection errors.

\subsubsection{Dataset Statistics}

The final dataset includes eight categories of reconnaissance scenarios as shown in Table~\ref{tab:categories}. Each category includes 129 multi-round interaction groups, and each group comprises three rounds of command-output pairs. The dataset contains 1{,}032 test cases in total.

\subsection{Testing Protocol}
For single-round evaluation, we use the first command of each case without prior interaction history. For multi-round evaluation, we issue all three commands in order. During LLM response generation, Rounds~2 and~3 receive the preceding commands and generated outputs from the same conversation. Reference outputs are never included in this history.
All three commands are constructed before evaluation and remain fixed across systems. Later responses therefore depend on the interaction history, but later commands are not rewritten from the generated outputs. An entity named in a follow-up command may have been specified during dataset construction. Appendix~\ref{command_example} presents a concrete three-round Linux
reconnaissance interaction, including the commands and reference outputs.

\subsection{Evaluation Metrics}

\textbf{Overview.} All three metrics in ReconPotBench are evaluated with an LLM-as-a-judge protocol. This choice is important because traditional metrics, such as exact matching or surface-level similarity, are too brittle for terminal-output evaluation: two outputs may differ in formatting while conveying the same system state, and a response may look locally plausible while still breaking entity continuity or offering little value for follow-up probing. Following our experimental pipeline, we use GPT-5.4-mini as a unified evaluator, and the full evaluation prompt is provided in Appendix~\ref{prompt_appendix}. This unified protocol ensures that Cowrie, closed-source LLMs, and open-source LLMs are assessed under the same standard. We evaluate each test subject from three perspectives: authenticity, cross-round consistency, and hacker-perspective utility, all computed on complete three-round reconnaissance interactions. These metrics are jointly necessary because they capture different properties of a LLM honeypot system. Authenticity measures realism with respect to real-system outputs, consistency measures whether entities are preserved, and utility measures whether the generated content remains valuable from the hacker's perspective. A system may perform well on one of these dimensions while performing poorly on another, so evaluating only a single metric would provide an incomplete benchmark.

Let $N$ denote the number of evaluated three-round reconnaissance
sequences, and let $\mathcal{I}_i$ denote sequence $i$.
We use the reference sequence in Appendix~\ref{command_example}
to illustrate the scoring rules. For this illustration, the
reference outputs themselves are treated as the evaluated outputs.
The illustrative labels are not additional experimental results
for any evaluated system.

\noindent\textbf{Authenticity Score.} Each command-output interaction receives one binary authenticity
score. For round $r$, let $c_r$, $o_r$, and $o_r^{\mathrm{ref}}$
denote the command, evaluated output, and corresponding reference
output, respectively. The evaluator judges agreement in command
semantics, the type of information returned, and terminal-output form:
\begin{equation}
\begin{gathered}
a_r=J_{\mathrm{auth}}
\bigl(c_r,o_r,o_r^{\mathrm{ref}}\bigr),
\\
a_r\in\{0,1\}.
\end{gathered}
\end{equation}
A score of 1 indicates agreement with the reference under these
criteria, and 0 indicates disagreement. This judgment does not
calculate string overlap.

The mean authenticity across the three rounds is
\begin{equation}
A=\frac{a_1+a_2+a_3}{3}.
\end{equation}

For the example in Appendix~\ref{command_example}, we illustrate
the calculation by treating the reference outputs themselves as
the evaluated outputs. Under the scoring rules, the illustrative
labels and their three-round mean are
\begin{equation}
\begin{gathered}
(a_1,a_2,a_3)=(1,1,1),
\\
A=\frac{1+1+1}{3}=1.
\end{gathered}
\end{equation}
These values illustrate the scoring rules and are not additional
experimental results for any evaluated system.

\noindent\textbf{Cross-Round Consistency Score.} Consistency is evaluated over each complete three-round sequence.
The evaluator assigns a label
$C_i \in \{0,1,\mathrm{NA}\}$ by examining entities and their
relationships across rounds. A label of 1 indicates consistency,
whereas 0 indicates a contradiction or a mismatch in these
relationships. Sequences without explicit entities are labeled
$\mathrm{NA}$.

Consistency checks whether entities are used consistently
across rounds. Different rounds may reveal different attributes
of the same entity, but its identity, previously reported
attribute values, and relationships with other entities must
not change without justification.

In the example in Appendix~\ref{command_example}, all three
rounds examine the same interface, \texttt{wlan0}.
Round~1 reports its addresses, Round~2 reports its network
configuration, and Round~3 reports its associated device
information. The interface identity is preserved across
rounds, with no conflicting attribute values or mismatched
entity references, giving the illustrative label
\begin{equation}
C_i = 1.
\end{equation}

Across the dataset, the consistency score is
\begin{equation}
\overline{C} = \frac{N_1}{N_0 + N_1},
\end{equation}
where $N_1$ and $N_0$ are the numbers of complete three-round
sequences labeled consistent and inconsistent, respectively.
Three-round sequences labeled $\mathrm{NA}$ are excluded
from the denominator.

\noindent\textbf{Attacker-Perspective Utility Score.} Utility is evaluated over each complete three-round sequence.
The evaluator assigns an integer score
$u_i \in \{0,1,2,3,4,5\}$ based on the outputs' relevance to
the commands, technical detail, and value for subsequent
reconnaissance. This score is assigned directly and is not
calculated from the authenticity or consistency scores.

Across the dataset, utility is aggregated as
\begin{equation}
\begin{gathered}
U = \sum_{s=0}^{5} s p_s,
\\
p_s = \frac{N_s}{N},
\end{gathered}
\end{equation}
where $N_s$ is the number of sequences assigned a utility
score of $s$, and $N$ is the total number of evaluated
sequences. Thus, $p_s$ is the observed proportion of sequences
with score $s$, and $U$ is the arithmetic mean of the
sequence-level utility scores.

Three-round sequences labeled $\mathrm{NA}$ are also included
in utility evaluation.

\noindent\textbf{Human Sampling Validation.} To validate the reliability of the evaluator, we manually inspect a balanced sample of 416 multi-round reconnaissance interactions, covering 40.31\% of all conversations. The sample is drawn from the Claude Sonnet-4.6 results, with 52 interactions selected from each of the eight reconnaissance categories. The validation was conducted by 2 graduate researchers, including 1 master's students and 1 Ph.D. students, and we compare manual judgments with GPT-5.4-mini outputs using the agreement-check template in Appendix~\ref{Agreement-Check}. Table~\ref{tab:judge-agreement} summarizes the agreement rates: 90.87\% for authenticity, 97.36\% for consistency, and 91.35\% for hacker-perspective utility. These results support agreement between human judgments and GPT-5.4-mini on the sampled Claude Sonnet-4.6 outputs.

\begin{table}[t]
\centering
\small
\begin{tabular}{lcc}
\toprule
\textbf{Metric} & \textbf{Matched / Total} & \textbf{Agreement} \\
\midrule
Authenticity & 1{,}134 / 1{,}248 & 90.87\% \\
Consistency & 405 / 416 & 97.36\% \\
Utility & 380 / 416 & 91.35\% \\
\bottomrule
\end{tabular}
\caption{\label{tab:judge-agreement}
Agreement between manual annotation and GPT-5.4-mini evaluator outputs on the sampled validation set.}
\end{table}

\section{Evaluation}
\label{sec:results}

We organize the evaluation around three research questions:

\begin{itemize}
    \item \textbf{RQ1:} What does jointly evaluating authenticity, cross-round consistency, and attacker-perspective utility reveal beyond the assessment of individual command outputs?

    \item \textbf{RQ2:} What trade-offs emerge among Cowrie, closed-source LLMs, and open-source LLMs across the three evaluation metrics?

    \item \textbf{RQ3:} How does system performance vary across reconnaissance categories in terms of authenticity, cross-round consistency, and attacker-perspective utility?
\end{itemize}

\subsection{Experimental Setup}
% 下面的表格是展示具体的评估分数
\begin{table*}[t]
  \centering
  \begin{tabular}{>{\raggedright\arraybackslash}m{0.26\textwidth}ccccccc}
    \hline
    \multirow{2}{*}{\textbf{System}}
    & \multicolumn{4}{c}{\textbf{Authenticity Score}}
    & \multirow{2}{*}{\shortstack[c]{\textbf{Consistency}\\\textbf{Score}}}
    & \multirow{2}{*}{\shortstack[c]{\textbf{Hacker Utility}\\\textbf{Score}}}
    & \multirow{2}{*}{\textbf{Rank}} \\
    \cline{2-5}
    & \textbf{R1} & \textbf{R2} & \textbf{R3} & \textbf{Avg.} & & & \\
    \hline

    \multicolumn{8}{l}{\textbf{Traditional Honeypot}} \\
    Cowrie
      & 0.21 & 0.15 & 0.15 & 0.17 & \ecomax{0.45} & 0.66 & 17 \\
    \hline

    \multicolumn{8}{l}{\textbf{Closed-source LLM}} \\
    Sonnet-4.6
      & \ecomax{0.36} & 0.33 & 0.19 & 0.29 & 0.11 & \ecomax{3.24} & \ecomax{1} \\
    Doubao-1.5
      & 0.26 & 0.21 & 0.09 & 0.19 & 0.08 & 2.70 & 13 \\
    Gemini-2.5
      & 0.31 & 0.24 & 0.09 & 0.21 & 0.10 & 2.88 & 8 \\
    GPT-5-nano
      & 0.33 & 0.27 & 0.13 & 0.24 & 0.12 & 3.04 & 7 \\
    Moonshot-v1
      & 0.30 & 0.19 & 0.10 & 0.20 & 0.12 & 2.79 & 9 \\
    GPT-5.1
      & 0.35 & 0.33 & \ecomax{0.22} & \ecomax{0.30} & 0.15 & 2.95 & 3 \\
    Gemini-3-Flash
      & 0.33 & \ecomax{0.35} & \ecomax{0.22} & \ecomax{0.30} & 0.16 & 2.94 & \ecomax{1} \\
    Qwen3-Max
      & 0.34 & 0.32 & 0.16 & 0.27 & 0.14 & 2.93 & 5 \\
    DeepSeek-V3.2
      & 0.33 & 0.27 & 0.16 & 0.25 & 0.12 & 3.02 & 6 \\
    DeepSeek-R1
      & 0.35 & 0.27 & 0.17 & 0.26 & 0.13 & 3.12 & 4 \\
    \hline

    \multicolumn{8}{l}{\textbf{Open-source LLM}} \\
    Hunyuan-7B
      & 0.01 & 0.05 & 0.03 & 0.03 & 0.12 & 1.18 & 22 \\
    Llama3.1-8B
      & 0.20 & 0.15 & 0.05 & 0.13 & 0.09 & 2.83 & 15 \\
    Mistral-7B
      & 0.18 & 0.13 & 0.06 & 0.12 & 0.07 & 2.56 & 19 \\
    Qwen2.5-7B
      & 0.28 & 0.25 & 0.10 & 0.21 & 0.12 & 2.38 & 14 \\
    Qwen3-0.6B (unthinking)
      & 0.04 & 0.05 & 0.03 & 0.04 & 0.19 & 0.72 & 23 \\
    Qwen3-0.6B (thinking)
      & 0.04 & 0.04 & 0.01 & 0.03 & 0.10 & 0.77 & 24 \\
    Qwen3-8B (unthinking)
      & 0.27 & 0.19 & 0.09 & 0.18 & 0.12 & 2.28 & 16 \\
    Qwen3-8B (thinking)
      & 0.21 & 0.16 & 0.08 & 0.15 & 0.13 & 2.35 & 18 \\
    Qwen3-14B (unthinking)
      & 0.26 & 0.23 & 0.10 & 0.20 & 0.13 & 2.64 & 10 \\
    Qwen3-14B (thinking)
      & 0.27 & 0.25 & 0.12 & 0.21 & 0.13 & 2.51 & 11 \\
    SecGPT-1.5B
      & 0.15 & 0.14 & 0.07 & 0.12 & 0.21 & 1.20 & 21 \\
    SecGPT-7B
      & 0.19 & 0.15 & 0.08 & 0.14 & 0.13 & 1.90 & 20 \\
    SecGPT-14B
      & 0.25 & 0.21 & 0.10 & 0.19 & 0.13 & 2.58 & 12 \\
    \hline
  \end{tabular}
  \caption{\label{tab:overall-evaluation-results}
    Comprehensive evaluation results for Cowrie and LLM honeypots in multi-round Linux reconnaissance. Authenticity is reported for each round, with ``Avg.'' denoting the mean across three rounds. Consistency is the fraction of conversations scored as fully consistent. Hacker utility is the weighted average of conversation-level scores from 0 to 5. Rank is computed by first converting hacker utility from the 0 to 5 scale to the 0 to 1 scale by dividing it by 5, and then averaging the converted hacker utility score with authenticity average and consistency. Higher values indicate better performance; all metric values are rounded to two decimals.
  }
\end{table*}

\textbf{Benchmark construction, response generation, and scoring.} We evaluate Cowrie~\citep{cowrie_github}, closed-source LLMs, and open-source LLMs on a multi-round Linux reconnaissance benchmark containing 1{,}032 interactions. The evaluation has three separate stages. During dataset construction, GPT-5.4 extends the manually designed first-round commands into three-round interactions, which are reviewed and paired with real-system reference outputs. Once the command set is fixed, running the benchmark does not require calling GPT-5.4 again. During response generation, the evaluated systems receive the fixed command sequences; later LLM responses also receive the preceding commands and generated outputs. Finally, GPT-5.4-mini scores the saved responses using the prompts in Appendix~\ref{prompt_appendix}. Reference outputs are used for authenticity scoring and are never supplied during response generation. Regenerating responses or rescoring them may still be affected by changes to commercial model services.

\noindent\textbf{Prompt control.} All evaluated LLMs receive the same system prompt, described in Appendix~\ref{Experimental_Details}. This holds the prompt text constant across models. The results characterize these models under the shared prompting setup and do not establish their best achievable performance under other prompts or state-management architectures.

\noindent\textbf{Test subjects and runtime.} We compare Cowrie with representative closed-source and open-source LLMs under the same benchmark. Closed-source LLMs are accessed through an OpenAI-compatible API pipeline, whereas open-source LLMs are executed locally. The closed-source LLMs evaluated in this work include Moonshot-v1-128k (Moonshot-v1), Doubao-1-5-lite-32k-250115 (Doubao-1.5), Gemini-2.5-flash-lite (Gemini-2.5), GPT-5-nano, Claude-sonnet-4-6 (Sonnet-4.6), GPT-5.1, Gemini-3-Flash, Qwen3-Max, DeepSeek-V3.2, and DeepSeek-R1. The open-source LLMs include Qwen2.5-7B-Instruct (Qwen2.5-7B), Llama-3.1-8B-Instruct (Llama3.1-8B), Mistral-7B-Instruct (Mistral-7B), Hunyuan-7B, Qwen3-0.6B/8B/14B with and without thinking mode, and SecGPT-1.5B/7B/14B. Detailed runtime environments and inference configurations are reported in Appendix~\ref{Experimental_Details}.

\noindent\textbf{Automatic evaluation.} We adopt GPT-5.4-mini to evaluate the multi-round outputs of each test subject from three perspectives. Authenticity for each round is determined by comparing the output of the test subject with the corresponding label output. Cross-round consistency is marked at the conversation level with values in \{0, 1, \texttt{NA}\}, based on whether key entities including IP addresses, ports, interface names, service names, process identifiers, routes, and paths stay consistent across rounds. The utility evaluated from a hacker perspective is scored per interaction as an integer from 0 to 5, which reflects how informative and valuable the generated content is for actual hackers' reconnaissance efforts.

\subsection{RQ1: Value of Joint Evaluation}
\label{subsec:rq1}

Table~\ref{tab:overall-evaluation-results} shows why individual-output authenticity is insufficient to characterize a reconnaissance interaction. GPT-5.1 and Gemini-3-Flash share the highest average authenticity score, at 0.30, but their consistency scores are only 0.15 and 0.16. Sonnet-4.6 achieves the highest attacker utility, at 3.24, while its consistency is 0.11. These system-level results distinguish the realism and information value of generated outputs from their compatibility across a conversation.

Authenticity is generally lower in later rounds, but the pattern is not universal. Sonnet-4.6 decreases from 0.36 to 0.33 to 0.19, whereas Hunyuan-7B increases from 0.01 to 0.05 before decreasing to 0.03.

\begin{challengebox}
\textbf{Finding 1.} Joint evaluation reveals weaknesses in cross-round consistency that individual-output authenticity does not capture. Systems with the highest authenticity or utility scores can still perform poorly on consistency.
\end{challengebox}

\begin{table*}[t]
  \centering
  \renewcommand{\arraystretch}{1.3}

  \begin{tabular}{@{}p{0.35\textwidth}cccccc@{}}
    \hline
    \textbf{Category} & \textbf{R1} & \textbf{R2} & \textbf{R3} & \textbf{Auth. Avg.} & \textbf{Consistency} & \textbf{Utility Avg.} \\
    \hline
    Network Interfaces \& IP & 0.32 & 0.26 & 0.16 & 0.25 & 0.04 & 3.17 \\
    Open Ports \& Listening Services & 0.49 & 0.36 & 0.22 & 0.36 & 0.18 & 3.40 \\
    Port Connection States & 0.26 & 0.28 & 0.35 & 0.30 & 0.20 & 3.64 \\
    Routing Table & 0.14 & 0.34 & 0.05 & 0.18 & 0.03 & 2.93 \\
    Service Versions \& Components & 0.60 & 0.32 & 0.16 & 0.36 & 0.08 & 3.35 \\
    DNS Configuration & 0.31 & 0.19 & 0.12 & 0.21 & 0.02 & 2.98 \\
    ARP Cache \& MAC Addresses & 0.24 & 0.64 & 0.16 & 0.35 & 0.01 & 2.84 \\
    Connection-to-Process Mapping & 0.55 & 0.26 & 0.26 & 0.36 & 0.30 & 3.57 \\
    \hline
  \end{tabular}
  \vspace{0.4em}
  \caption{Category-level evaluation scores for Claude Sonnet-4.6. The 1,032 interactions are ordered as eight blocks of 129 interactions following Table~\ref{tab:categories}. R1 to R3 report round-wise authenticity scores, ``Auth. Avg.'' reports the average authenticity across the three rounds, ``Consistency'' reports the fraction of conversations with consistency score 1, and ``Utility Avg.'' reports the weighted average hacker-utility score. All reported values are rounded to two decimal places.}
  \label{tab:sonnet46-category-results}
\end{table*}

\subsection{RQ2: Comparison Across Systems}
\label{subsec:rq2}

Cowrie achieves the highest consistency score, at 0.45, compared with 0.07 to 0.21 across the evaluated LLM configurations. Cowrie's constrained command support and responses may contribute to this difference.

The authenticity comparison varies by model. All evaluated closed-source models exceed Cowrie's average authenticity of 0.17. Open-source results are more mixed: Qwen2.5-7B scores 0.21, while Llama3.1-8B and Hunyuan-7B score 0.13 and 0.03, respectively. Therefore, the results do not support a general claim that open-source LLMs outperform Cowrie on authenticity.

Every evaluated LLM configuration exceeds Cowrie on attacker utility. Cowrie scores 0.66, compared with mean scores of 1.99 and 2.96 across the evaluated open-source and closed-source configurations. This comparison concerns the reconnaissance information supplied by the outputs.

The paired Qwen3 results also show no consistent benefit from enabling thinking mode. Depending on model size, improvements in one metric accompany decreases or little change in another. Appendix~\ref{Thinking_Mode_Analysis} reports these comparisons in detail.

We test all LLMs with the same prompt and do not include a baseline that explicitly stores and updates the simulated host’s state. The results therefore describe performance under this setup. They do not establish the best performance these models could achieve with different prompts or ways of maintaining that state.

\begin{challengebox}
\textbf{Finding 2.} The evaluated LLMs all provide higher reconnaissance utility than Cowrie, but authenticity gains depend on the model, and none matches Cowrie's consistency. No evaluated system leads across all three metrics.
\end{challengebox}

\subsection{RQ3: Category-Level Performance}
\label{subsec:rq3}

We examine category-level performance using the Claude Sonnet-4.6 results in Table~\ref{tab:sonnet46-category-results}, comparing its authenticity, cross-round consistency, and attacker-perspective utility across the eight reconnaissance categories. Consistency is highest for Connection-to-Process Mapping (0.30), followed by Port Connection States (0.20) and Open Ports \& Listening Services (0.18). It is lower for Network Interfaces \& IP (0.04), Routing Table (0.03), DNS Configuration (0.02), and ARP Cache \& MAC Addresses (0.01).

Similar authenticity scores can accompany different consistency scores. Connection-to-Process Mapping, Open Ports \& Listening Services, and Service Versions \& Components all have average authenticity 0.36, but their consistency scores are 0.30, 0.18, and 0.08. Port Connection States has the highest utility, at 3.64, while Connection-to-Process Mapping has the highest consistency. ARP Cache \& MAC Addresses further illustrates the separation between metrics, with average authenticity 0.35 and consistency 0.01.

Round-level authenticity is also nonmonotonic in several categories. Port Connection States increases from 0.26 to 0.28 to 0.35, while ARP Cache \& MAC Addresses increases from 0.24 to 0.64 before falling to 0.16. These results identify differences among the selected tasks.

\begin{challengebox}
\textbf{Finding 3.} For Claude Sonnet-4.6, performance varies across reconnaissance categories and across the three metrics. Categories with similar authenticity can have substantially different cross-round consistency, and the category with the highest attacker utility is not the most consistent.
\end{challengebox}

\section{Conclusion}
\label{sec:conclusion}

We present ReconPotBench, a benchmark for evaluating LLM honeypots in multi-round Linux reconnaissance. It contains 1{,}032 three-round interactions across eight command categories, with real-system reference outputs, and evaluates systems along authenticity, cross-round consistency, and hacker-perspective utility. ReconPotBench reveals weaknesses in cross-round consistency that individual-output authenticity does not capture. Across systems, every evaluated LLM configuration provides higher reconnaissance utility than Cowrie, but authenticity gains depend on the model and Cowrie achieves the highest consistency. Within Claude Sonnet-4.6, similar authenticity scores can accompany substantially different consistency across categories.

% Use \verb|\appendix| before any appendix section to switch the section numbering over to letters. See Appendix~\ref{sec:appendix} for an example.

\section*{Limitations}
ReconPotBench is designed to evaluate LLM honeypots in multi-round Linux reconnaissance. Its three-round structure allows us to collect reliable real-system references and compare different systems under a reproducible setting. However, the current benchmark focuses on single-goal reconnaissance chains within Linux terminal environments, and does not yet cover cross-goal reconnaissance behavior or other operating systems and shell environments. Future work can extend ReconPotBench to broader platforms and more complex multi-stage reconnaissance workflows.

\section*{Ethical Considerations}

All data in ReconPotBench is collected in a controlled local environment. Commands are executed only on our own physical Linux host, and we do not probe third-party websites, domains, or servers. Thus, our data collection does not affect the security or privacy of external systems or individuals. ReconPotBench may have dual-use implications, since methods for evaluating honeypots could be misused by attackers to better recognize or avoid honeypots. We release the benchmark only for academic and defensive research, with the goal of supporting reproducible evaluation and improving the robustness of LLM-based honeypots.

\section*{Acknowledgments}

This work was supported in part by the Science and Technology Development Fund of Macau SAR (Grant Nos. FDCT/0007/2024/AKP, EF2024-00185-FST), the UM and UMDF (Grant Nos. MYRG-GRG2025-00236-FST, MYRG-GRG2026-00018-FIC, MYRG-CRG2026-00007-FIC), the Dr. Stanley Ho Medical Development Foundation (Grant No. SHMDF-AI/2026/001), and the National Natural Science Foundation of China (Grant No. 62266013).

% This document does not cover the content requirements for ACL or any
% other specific venue.  Check the author instructions for
% information on
% maximum page lengths, the required ``Limitations'' section,
% and so on.

% \section*{Acknowledgments}

% This document has been adapted
% by Steven Bethard, Ryan Cotterell and Rui Yan
% from the instructions for earlier ACL and NAACL proceedings, including those for
% ACL 2019 by Douwe Kiela and Ivan Vuli\'{c},
% NAACL 2019 by Stephanie Lukin and Alla Roskovskaya,
% ACL 2018 by Shay Cohen, Kevin Gimpel, and Wei Lu,
% NAACL 2018 by Margaret Mitchell and Stephanie Lukin,
% Bib\TeX{} suggestions for (NA)ACL 2017/2018 from Jason Eisner,
% ACL 2017 by Dan Gildea and Min-Yen Kan,
% NAACL 2017 by Margaret Mitchell,
% ACL 2012 by Maggie Li and Michael White,
% ACL 2010 by Jing-Shin Chang and Philipp Koehn,
% ACL 2008 by Johanna D. Moore, Simone Teufel, James Allan, and Sadaoki Furui,
% ACL 2005 by Hwee Tou Ng and Kemal Oflazer,
% ACL 2002 by Eugene Charniak and Dekang Lin,
% and earlier ACL and EACL formats written by several people, including
% John Chen, Henry S. Thompson and Donald Walker.
% Additional elements were taken from the formatting instructions of the \emph{International Joint Conference on Artificial Intelligence} and the \emph{Conference on Computer Vision and Pattern Recognition}.

\bibliography{custom}

\appendix

\section{Prompt Paradigm}
\label{prompt_appendix}

\paragraph{Prompt for Authenticity Judge.}
\label{Authenticity-judge}

\begin{tcolorbox}[
  title={Prompt for Authenticity Evaluation},
  breakable,
  fonttitle=\bfseries,
  enhanced,
  colback=generate!10,
  colbacktitle=generate,
  coltitle=black,
  colframe=generate!80!black,
  coltext=black,
  boxrule=0.5pt,
  arc=2mm
]
\small

\textbf{Role.}
You are an evaluator of Linux terminal output authenticity.

\textbf{Task.}
Your task is to evaluate authenticity at the single-round level within a multi-round interaction. For each evaluation instance, consider only the current round and determine whether the LLM-generated output is similar to the corresponding real system output for the same Linux reconnaissance command.

\textbf{Input.}
\texttt{input\_cmd}: the Linux reconnaissance command at the current round.\\
\texttt{test\_output}: the output generated by the LLM for this command.\\
\texttt{system\_output}: the real system output for the same command, used as the reference output.

\textbf{Evaluation Criterion.}
Judge whether \texttt{test\_output} aligns with \texttt{system\_output} in terms of the semantics, the type of returned information, and the overall form of the terminal output for \texttt{input\_cmd}.

\textbf{Scoring.}
Output 1 if \texttt{test\_output} aligns with \texttt{system\_output}.\\
Output 0 if \texttt{test\_output} does not align with \texttt{system\_output}.

\textbf{Output Requirement.}
Return exactly one character: 1 or 0. Do not output anything else.
\end{tcolorbox}

\paragraph{Prompt for Hacker-Perspective Judge.}
\label{Hacker-Perspective}

\begin{tcolorbox}[
  title={Prompt for Hacker Utility Evaluation},
  breakable,
  fonttitle=\bfseries,
  enhanced,
  colback=generate!10,
  colbacktitle=generate,
  coltitle=black,
  colframe=generate!80!black,
  coltext=black,
  boxrule=0.5pt,
  arc=2mm
]
\small

\textbf{Role.}
You are an evaluator of multi-round Linux reconnaissance conversations from a hacker-perspective utility standpoint.

\textbf{Task.}
Your task is to determine whether the conversation outputs, taken together, provide reconnaissance information that is useful and meaningful from a hacker’s perspective.

\textbf{Input.}
The input unit is one complete ``conversation'' from XXX file, which contains 3 rounds.

For each round, represent the conversation as a command-output pair in the following format:
\begin{itemize}
    \item \texttt{contents}: the concatenated string of all prior rounds up to the current command, ending with the current command
    \item \texttt{output}: the corresponding output for that current round
\end{itemize}

Specifically, the 3 rounds should be interpreted as:

\textit{Round 1.}
\begin{itemize}
    \item \texttt{contents} = Round 1 command
    \item \texttt{output} = Round 1 output
\end{itemize}

\textit{Round 2.}
\begin{itemize}
    \item \texttt{contents} = Round 1 command + Round 1 output + Round 2 command
    \item \texttt{output} = Round 2 output
\end{itemize}

\textit{Round 3.}
\begin{itemize}
    \item \texttt{contents} = Round 1 command + Round 1 output + Round 2 command + Round 2 output + Round 3 command
    \item \texttt{output} = Round 3 output
\end{itemize}

You must evaluate the full 3-round conversation as one unit, based on the progression across these command-output pairs.

\textbf{Evaluation Criteria.}
Assign an integer score from \texttt{0} to \texttt{5} according to the following standard:
\begin{itemize}
    \item \texttt{0}: The outputs are entirely useless from a hacker’s perspective. They are empty, irrelevant, obviously mismatched to the commands, or provide no meaningful reconnaissance value.
    \item \texttt{1}: The outputs contain only minimal or fragmentary reconnaissance value. Most information is vague, generic, poorly matched to the commands, or not useful for guiding further investigation.
    \item \texttt{2}: The outputs provide limited reconnaissance value. Some command-related information is present, but it is shallow, incomplete, weakly informative, or insufficient to meaningfully support further reconnaissance.
    \item \texttt{3}: The outputs provide moderate reconnaissance value. They are generally relevant to the commands and contain some useful technical details, but the information is not consistently specific, well-connected, or strong enough to form a highly useful reconnaissance chain.
    \item \texttt{4}: The outputs provide strong reconnaissance value. They are relevant to the commands, contain substantial attacker-relevant technical details, and show a mostly meaningful progression across the three rounds. However, at least one noticeable limitation remains, such as incomplete specificity, weaker progression between some rounds, partial genericity, or missing details that would prevent the conversation from being maximally useful for guiding the next reconnaissance step.
    \item \texttt{5}: The outputs provide very strong reconnaissance value. They are consistently relevant to the commands, rich in concrete attacker-relevant technical details, and form a clear and coherent reconnaissance chain across all three rounds. The outputs not only reveal useful information, but also provide enough specific, well-connected, and operationally actionable detail to directly support plausible next-step attacker decision-making, with no major weakness in relevance, specificity, or progression.
\end{itemize}

\textbf{Output Requirement.}
Return only one integer: \texttt{0}, \texttt{1}, \texttt{2}, \texttt{3}, \texttt{4}, or \texttt{5}.
\end{tcolorbox}

\paragraph{Prompt for Consistency Judge.}
\label{sec:appendix_prompt}
\begin{tcolorbox}[
  title={Prompt for Consistency Evaluation},
  breakable,
  fonttitle=\bfseries,
  enhanced,
  colback=generate!10,
  colbacktitle=generate,
  coltitle=black,
  colframe=generate!80!black,
  coltext=black,
  boxrule=0.5pt,
  arc=2mm
]
\small

\textbf{Role.}
You are an evaluator of multi-round Linux reconnaissance dialogues from a cross-round consistency standpoint.

\textbf{Task.}
Your task is to determine whether the conversation maintains full contextual consistency across rounds when entities are present.

\textbf{Input.}
The input unit is one complete ``conversation'' from XXX file, which contains 3 rounds.

For each round, represent the conversation as a command-output pair in the following format:
\begin{itemize}[leftmargin=1.5em, itemsep=0.2em, topsep=0.2em]
    \item \texttt{contents}: the concatenated string of all prior rounds up to the current command, ending with the current command
    \item \texttt{output}: the corresponding output for that current round
\end{itemize}

Specifically, the 3 rounds should be interpreted as:

\textit{Round 1.}
\begin{itemize}[leftmargin=1.5em, itemsep=0.2em, topsep=0.2em]
    \item \texttt{contents} = Round 1 command
    \item \texttt{output} = Round 1 output
\end{itemize}

\textit{Round 2.}
\begin{itemize}[leftmargin=1.5em, itemsep=0.2em, topsep=0.2em]
    \item \texttt{contents} = Round 1 command + Round 1 output + Round 2 command
    \item \texttt{output} = Round 2 output
\end{itemize}

\textit{Round 3.}
\begin{itemize}[leftmargin=1.5em, itemsep=0.2em, topsep=0.2em]
    \item \texttt{contents} = Round 1 command + Round 1 output + Round 2 command + Round 2 output + Round 3 command
    \item \texttt{output} = Round 3 output
\end{itemize}

You must evaluate the full 3-round conversation as one unit, based on the progression across these command-output pairs.

\textbf{Evaluation Criteria.}
Judge the dialogue according to the following criteria:
\begin{enumerate}[leftmargin=1.5em, itemsep=0.2em, topsep=0.2em]
    \item First determine whether the multi-round interaction contains explicit entities, such as IP addresses, ports, interface names, service names, process identifiers, routes, paths, or device names.
    \item If no such entities appear anywhere in the conversation, do not score this conversation for consistency.
    \item If entities do appear, check whether they are used consistently across all rounds.
    \item Entity consistency means that the same entity, once introduced, should not be changed, contradicted, mismatched, or replaced without justification in later rounds.
    \item Cross-round context must remain fully aligned between commands and outputs. Any inconsistency in entity identity, value, reference, or contextual linkage counts as inconsistency.
    \item Judge only contextual consistency across rounds.
\end{enumerate}

\textbf{Scoring.}
\begin{itemize}[leftmargin=1.5em, itemsep=0.2em, topsep=0.2em]
    \item Output \textbf{1} if the conversation contains entities and the context is fully consistent across rounds.
    \item Output \textbf{0} if the conversation contains entities but the context is inconsistent across rounds.
    \item Output \textbf{NA} if the conversation does not contain any entities and therefore should be excluded from the average consistency calculation.
\end{itemize}

\textbf{Output Requirement.}
Return only a single value: 1, 0, or NA.

\end{tcolorbox}

\paragraph{Prompt for Round 2/3 Command Generation.}
\label{para:appendix_multiround_prompt}
\begin{tcolorbox}[
  title={Round 2/3 Command Prompt},
  breakable,
  fonttitle=\bfseries,
  enhanced,
  colback=generate!10,
  colbacktitle=generate,
  coltitle=black,
  colframe=generate!80!black,
  coltext=black,
  boxrule=0.5pt,
  arc=2mm
]
\small
\textbf{Role:} You are a Linux reconnaissance command generation assistant.

\textbf{Task:} Given a user-provided Round 1 command for network-layer reconnaissance, generate exactly two follow-up commands: one for Round 2 software-layer reconnaissance and one for Round 3 hardware-layer reconnaissance. The three rounds must strictly follow the progression Network $\rightarrow$ Software $\rightarrow$ Hardware.

\textbf{Rules:} First infer which of the following eight reconnaissance scenarios the Round 1 command belongs to: Network Interfaces and IP, Open Ports and Listening Services, Network Connections and Port States, Routing Table, Service Versions and Components, DNS Configuration, ARP Cache and MAC Addresses, or Connection-to-Process Mapping. Then generate Round 2 and Round 3 so that both remain within the same scenario and preserve logical dependence across rounds. The Round 2 command must target software information motivated by Round 1, and the Round 3 command must target hardware information motivated by Round 2. The generated commands must be valid, commonly used, non-destructive Linux commands. When applicable, context-dependent entities such as IP addresses, ports, interface names, service names, process identifiers, routes, paths, and device names should remain consistent across rounds. If the input does not contain a concrete entity value, generate a general but still scenario-consistent follow-up command. Do not modify or repeat the user-provided Round 1 command.

\textbf{Input:} \texttt{first\_command: \{user\_provided\_round1\_command\}}

\textbf{Output:} Exactly two lines only. Line 1 is the Round 2 command. Line 2 is the Round 3 command. Do not output any explanation, label, numbering, or additional text.

\textbf{Examples:}

\texttt{Example 1:} \\
\texttt{Input: the command does inspect network interfaces, IP addresses, and host-side addressing configuration.} \\
\texttt{Output: Software layer: the command does inspect software or configuration responsible for interface management and host network settings.} \\
\texttt{Hardware layer: the command does inspect the network interface hardware that supports those interfaces and addresses.}

\texttt{Example 2:} \\
\texttt{Input: the command does inspect open ports and listening services exposed by the host.} \\
\texttt{Output: Software layer: the command does inspect the service software or service configuration corresponding to those listening ports.} \\
\texttt{Hardware layer: the command does inspect the hardware resources or network devices supporting those exposed services.}

\texttt{Example 3:} \\
\texttt{Input: the command does inspect active connections, connection states, and current communication status.} \\
\texttt{Output: Software layer: the command does inspect the software processes or service-side software state associated with those active connections.} \\
\texttt{Hardware layer: the command does inspect the hardware resources supporting the current connection workload.}

\texttt{Example 4:} \\
\texttt{Input: the command does inspect routing-table information and forwarding paths used for network reachability.} \\
\texttt{Output: Software layer: the command does inspect routing-related or firewall-related software and configuration.} \\
\texttt{Hardware layer: the command does inspect the network hardware supporting packet forwarding or route-dependent communication.}

\texttt{Example 5:} \\
\texttt{Input: the command does inspect service versions and exposed service components.} \\
\texttt{Output: Software layer: the command does inspect installed packages and configuration associated with the identified software version.} \\
\texttt{Hardware layer: the command does inspect platform or hardware characteristics that constrain software compatibility or deployment.}

\texttt{Example 6:} \\
\texttt{Input: the command does inspect DNS configuration and resolver-related network settings.} \\
\texttt{Output: Software layer: the command does inspect DNS-related software, resolver configuration, or name-service components.} \\
\texttt{Hardware layer: the command does inspect the hardware resources supporting DNS-related operation.}

\texttt{Example 7:} \\
\texttt{Input: the command does inspect ARP cache entries, MAC addresses, and neighbor information at the link layer.} \\
\texttt{Output: Software layer: the command does inspect software or configuration related to neighbor discovery or local network interface management.} \\
\texttt{Hardware layer: the command does inspect the network interface hardware involved in link-layer communication.}

\texttt{Example 8:} \\
\texttt{Input: the command does inspect the mapping between network connections and processes.} \\
\texttt{Output: Software layer: the command does inspect the binaries, libraries, configuration, or process-level software components associated with the identified process or service.} \\
\texttt{Hardware layer: the command does inspect the hardware resources supporting that process, such as CPU, memory, or relevant devices.}
\end{tcolorbox}

\section{Human Agreement Check Template}
\label{Agreement-Check}
\begin{tcolorbox}[
  title={Template for Human Agreement Check},
  breakable,
  fonttitle=\bfseries,
  enhanced,
  colback=generate!10,
  colbacktitle=generate,
  coltitle=black,
  colframe=generate!80!black,
  coltext=black,
  boxrule=0.5pt,
  arc=2mm
]
\small

\textbf{Task.}
For each sampled validation item, independently assign a human judgment using the same scoring rule as the corresponding evaluator prompt. Then compare the human judgment with the GPT-5.4-mini judgment.

\textbf{Input.}
\texttt{metric\_type}: authenticity, consistency, or hacker-perspective utility.\\
\texttt{sample\_content}: the complete three-round interaction to be judged.\\
\texttt{reference\_output}: the real-system output, used only for authenticity judgments.\\
\texttt{gpt\_judge\_output}: the score produced by GPT-5.4-mini.

\textbf{Agreement Rule.}
\begin{itemize}[leftmargin=1.5em, itemsep=0.2em, topsep=0.2em]
    \item For authenticity, mark agreement if the human label and GPT-5.4-mini label are both 1 or both 0.
    \item For consistency, mark agreement if the human label and GPT-5.4-mini label are both 1, both 0, or both \texttt{NA}.
    \item For hacker utility, mark agreement if the human score and GPT-5.4-mini score are the same integer from 0 to 5.
\end{itemize}

\end{tcolorbox}

\onecolumn
\section{Concrete Command Example}
\label{command_example}

\begin{challengebox}
\normalfont\small
\textbf{Example: A fixed three-round reconnaissance interaction.}
\par\smallskip

This case from the released dataset examines the same network
interface, \texttt{wlan0}, through its network state, software
configuration, and underlying device information. The outputs below
are collected from the reference host.

\par\medskip
\noindent\textbf{Round 1: Network state and addresses.}

{\footnotesize
\begin{verbatim}
Command:
ip -br addr show dev wlan0

Reference output:
wlan0            UP             192.168.0.100/24 fe80::ca35:41ed:8341:c564/64
\end{verbatim}
}

The output identifies \texttt{wlan0} as an active interface with an
IPv4 address and an IPv6 link-local address. It establishes the
interface and its network state for this case.

\par\medskip
\noindent\textbf{Round 2: Software configuration of the same interface.}

{\footnotesize
\begin{verbatim}
Command:
sysctl -a 2>/dev/null | grep -E '^net\.(ipv4|ipv6)\.(conf|neigh)\.wlan0(\.|$)' | sed -n '1,5p'

Reference output:
net.ipv4.conf.wlan0.accept_local = 0
net.ipv4.conf.wlan0.accept_redirects = 1
net.ipv4.conf.wlan0.accept_source_route = 0
net.ipv4.conf.wlan0.arp_accept = 0
net.ipv4.conf.wlan0.arp_announce = 0
\end{verbatim}
}

This round adds kernel network settings for the interface.
Every returned entry belongs to \texttt{net.ipv4.conf.wlan0},
so the configuration must refer to the same interface examined in
Round~1. These values provide configuration information; they do
not, by themselves, establish a vulnerability.

\par\medskip
\noindent\textbf{Round 3: Device context of the interface.}

{\footnotesize
\begin{verbatim}
Command:
ethtool -i wlan0 2>/dev/null | sed -n '1,5p'

Reference output:
driver: iwlwifi
version: 6.12.25-amd64
firmware-version: 89.4d42c933.0 so-a0-hr-b0-89.uc
expansion-rom-version:
bus-info: 0000:00:14.3
\end{verbatim}
}

The final round adds the driver, firmware, and bus location,
connecting the logical interface to its underlying device context.
This is the hardware-related information examined in this case.

\end{challengebox}
\twocolumn

\clearpage
\section{Experimental Details}
\label{Experimental_Details}

All LLM honeypots are evaluated with the same system prompt. The prompt instructs the LLM to act as a high-interaction honeypot agent operating as a realistic Kali Linux system. Given a reconnaissance command, the LLM is required to return terminal-style output that follows standard Linux conventions, reflects a believable Kali runtime environment, and provides valuable reconnaissance information. The prompt also requires the LLM to preserve details such as host configuration, network information, running services, file-system state, and active processes, and other relevant system artifacts across interaction rounds, so that the generated outputs support a continuous and informative reconnaissance workflow.

Table~\ref{tab:runtime-settings} reports the runtime environment and key inference configurations for open-source LLM generation.
\begin{table}[H]
  \centering
  \small
  \setlength{\tabcolsep}{8pt}
  \renewcommand{\arraystretch}{1.08}
  \begin{tabular}{ll}
    \hline
    \textbf{Configuration} & \textbf{Value} \\
    \hline
    Inference framework & \texttt{transformers} + \texttt{PyTorch} \\
    Sampling strategy & \texttt{do\_sample=False} \\
    Maximum generation length & \texttt{max\_new\_tokens=1024} \\
    GPU & NVIDIA A100-PCIE-40GB \\
    CPU & 10 Xeon vCPUs \\
    Operating system & Ubuntu 22.04 \\
    Python version & Python 3.12 \\
    PyTorch version & PyTorch 2.5.1 \\
    CUDA version & CUDA 12.4 \\
    \hline
  \end{tabular}
  \caption{\label{tab:runtime-settings}
    Runtime environment and inference settings for open-source LLM generation.
  }
\end{table}

\section{Thinking Mode Analysis}
\label{Thinking_Mode_Analysis}

We further examine whether enabling thinking mode improves multi-round honeypot performance within the Qwen3 family. The paired results show no consistent advantage. For Qwen3-0.6B, thinking mode slightly improves hacker-perspective utility from 0.72 to 0.77, but average authenticity decreases from 0.04 to 0.03, and consistency drops more clearly from 0.19 to 0.10. For Qwen3-8B, thinking mode reduces average authenticity from 0.18 to 0.15, while consistency increases only marginally from 0.12 to 0.13 and hacker-perspective utility from 2.28 to 2.35. For Qwen3-14B, thinking mode brings only a small authenticity gain, from 0.20 to 0.21, leaves consistency unchanged at 0.13, and lowers hacker-perspective utility from 2.64 to 2.51.

These results show no consistent benefit from enabling thinking mode under the tested settings. They do not isolate whether reasoning behavior, state management, or another factor accounts for the observed differences.

\clearpage
\twocolumn[
\begin{center}
  \renewcommand{\arraystretch}{1.3}
  \begin{tabular}{@{}p{0.04\textwidth}p{0.22\textwidth}p{0.34\textwidth}p{0.34\textwidth}@{}}
    \toprule
    \textbf{ID} &
    \shortstack{\textbf{Round 1}\\\textbf{Network-layer}} &
    \shortstack{\textbf{Round 2}\\\textbf{Software-layer}} &
    \shortstack{\textbf{Round 3}\\\textbf{Hardware-layer}} \\
    \midrule
    1 &
    Network Interfaces \& IP &
    Interface count, address type, and bound services &
    interface-related hardware \\

    2 &
    Open Ports \& Listening Services &
    Port-to-service mapping and version clues &
    Service-specific hardware needs \\

    3 &
    Port Connection States &
    Client/server software behind active connections &
    Load-related hardware bottlenecks \\

    4 &
    Routing Table &
    Gateway/subnet clues for routing or firewall software &
    Routing and NIC hardware support \\

    5 &
    Service Versions \& Components &
    Package, component, and vulnerability-related software clues &
    Deployment-related hardware support \\

    6 &
    DNS Configuration &
    DNS software stack &
    Local network role \\

    7 &
    ARP Cache \& MAC Addresses &
    ARP-related system utilities &
    LAN device identity and hardware details \\

    8 &
    Connection-to-Process Mapping &
    Process paths and software dependencies &
    Host resource and hardware context \\
    \bottomrule
  \end{tabular}
  \vspace{0.5em}
\captionof{table}{Eight reconnaissance categories and their three-round dependency chain.}
\label{tab:categories}
\end{center}
]

\section{Analysis of Reconnaissance Categories}
\label{Reconnaissance_Categories}

All eight scenarios are grounded in relevant reconnaissance references. Specifically, Network Interfaces \& IP is supported by~\citep{secjuice_recon,appati_network_recon}; Open Ports \& Listening Services is supported by~\citep{marceiano_enumeration,secjuice_recon}; Port Connection States corresponds to network connections and port-state reconnaissance in~\citep{secjuice_recon}; Routing Table is supported by~\citep{appati_network_recon}; Service Versions corresponds to port-service version reconnaissance in~\citep{marceiano_enumeration,appati_network_recon}; DNS Configuration corresponds to DNS reconnaissance in~\citep{appati_network_recon}; ARP Cache \& MAC Addresses corresponds to ARP reconnaissance in~\citep{ballaco_arp_scan}; and Connection-to-Process Mapping corresponds to network connections and their associated processes in~\citep{mitre_t1049}.

Table~\ref{tab:categories} lists all eight categories with representative commands. Table~\ref{tab:sonnet46-category-results} reports the category-level scores of Claude Sonnet-4.6 across the eight reconnaissance command families, including round-wise authenticity, average authenticity, cross-round consistency, and hacker-perspective utility.

\end{document}